\documentclass[letter]{aa} 
\usepackage{graphicx}
\usepackage{txfonts}
 \begin{document}

   \title{Dissipative and nonaxisymmetric standard-MRI in Kepler disks}

   \author{L.\,L.~Kitchatinov\inst{1,2,3} \and G.~R\"udiger\inst{1}
          }

   \institute{Astrophysikalisches Institut Potsdam, An der Sternwarte 16,
              D-14482, Potsdam, Germany \\
              \email{lkitchatinov@aip.de; gruediger@aip.de}
         \and
             Institute for Solar-Terrestrial Physics, PO Box
             291, Irkutsk 664033, Russia \\
             \email{kit@iszf.irk.ru}
         \and
             Pulkovo Astronomical Observatory, St. Petersburg, 196140, Russia
             }

   \date{Received ; accepted }

  \abstract{} 
{Deviations from axial symmetry are necessary to maintain self-sustained MRI-turbulence. We define the parameters region where nonaxisymmetric MRI is excited and study dependence of the unstable modes structure and growth rates on the relevant parameters.} 
{We solve numerically the linear eigenvalue problem for global axisymmetric and nonaxisymmetric modes of standard-MRI in Keplerian disks with finite diffusion.} 
{For  small magnetic Prandtl number the microscopic viscosity  completely drops out from the analysis so that the stability maps and the growth rates expressed in terms of the magnetic Reynolds number Rm and the Lundquist number S do not depend on the magnetic Prandtl number Pm. The minimum magnetic field for onset of nonaxisymmetric MRI grows with Rm. For given S all nonaxisymmetric modes disappear for sufficiently high Rm. This behavior is a consequence of the radial fine-structure of the nonaxisymmetric modes resulting from the winding effect of differential rotation. It is this fine-structure which presents severe resolution problems for the numerical simulation of MRI at large Rm.}
{For weak supercritical magnetic fields only axisymmetric modes are unstable. Nonaxisymmetric modes need stronger fields and not too fast rotation. If Pm is small its real value does not play any role in MRI.}
  

 \authorrunning{L.\,L.~Kitchatinov \& G.~R\"udiger}
 \titlerunning{Global nonaxisymmetric MRI}
\maketitle
\section{Introduction}
The leading concept for the origin of turbulence in accretion disks is the turbulence driving by the magnetorotational instability (MRI). The instability can be excited by even a very weak magnetic field provided that rotation with outward decreasing angular velocity is present (see Balbus \& Hawley \cite{BH98}).

MRI is expected to possess the  remarkable property of being self-sustained, i.e. to support the destabilizing magnetic field via its own dynamo (Brandenburg et al. \cite{BNST95}; Hawley et al. \cite{HGB96}).  The MRI-driven dynamo needs a finite but small initial field to start.  Differential rotation alone cannot drive a dynamo (Elsasser \cite{E46}). Deviations from axial symmetry are necessary for dynamo and the deviations have to be produced by the MRI itself. The excitation of nonaxisymmetric modes of MRI is thus necessary for the self-sustained turbulence.   

The present paper focuses on the nonaxisymmetric MRI. A model of differentially rotating disk with finite diffusivities and axial background field is applied to perform linear analysis of global stability. As in the axisymmetric case, the nonaxisymmetric MRI can be found in a range between some minimum $B_\mathrm{min}$ and maximum $B_\mathrm{max}$ values of the background field. The instability range depends on rotation rate (parameterized by the magnetic Reynolds number Rm). In contrast to the axisymmetric case, $B_\mathrm{min}$ for nonaxisymmetric modes does not approach a (small) constant value with increasing rotation but grows in linear proportion to Rm. The larger Rm, the stronger background field is required to maintain the  nonaxisymmetric instability.

This behavior is increasingly difficult to follow numerically as Rm grows because of the winding effect of differential rotation. The shearing of nonaxisymmetric fields by differential rotation produces fine radial structure when the field is too weak to resist the winding. This shearing effect can simultaneously be the reason for the increase of $B_\mathrm{min}$ with Rm and for the high resolution required to resolve the nonaxisymmetric MRI numerically (Fromang \& Papaloizou \cite{FP07}).
\section{The model}
We use the model of  Kitchatinov \& Mazur (\cite{KM97}) considering a rotating disk of constant thickness, $2H$, threaded by a uniform axial magnetic field. The rotation axis is normal to the disk plane and the angular velocity, $\Omega$, varies only with the distance $s$ to the axis 
\begin{equation}
  \Omega(s) = \Omega_0\left( 1 +\left(\frac{s}{s_0}\right)^{3}\right)^{-1/2} . 
  \label{1}
\end{equation}
This profile describes almost uniform rotation with the velocity $\Omega_0$ up to the distance $s_0$ and Kepler rotation  for larger distances.  The  aspect ratio is  $s_0/H = 5$.  
 We assume incompressibility, $\mathrm{div}\ \vec{U} = 0$. The motion equation is curled to exclude the pressure. This  results in the equation for vorticity, ${\vec W} = {\vec\nabla}\times{\vec U}$, 
\begin{equation}
  \frac{\partial{\vec W}}{\partial t} = 
  {\vec\nabla}\times\left( {\vec U}\times{\vec W} + 
  {\vec J}\times{\vec B}/\rho\right) + \nu\Delta{\vec W}, 
  \label{2}
\end{equation}
where ${\vec B}$ is magnetic field, ${\vec J} = \mu_0^{-1}{\vec\nabla}\times{\vec B}$ is the current density, and $\nu$ is the viscosity. The magnetic field follows the induction equation 
\begin{equation}
  \frac{\partial{\vec B}}{\partial t} = {\vec\nabla}\times
  \left( {\vec U}\times{\vec B} - \eta{\vec\nabla}\times{\vec B}\right) .
  \label{3}
\end{equation}
The equations are linearized about the background state of rotational motion (\ref{1}) and uniform axial magnetic field ${\vec B}_0 = \hat{\vec z} B_0$, where $\hat{\vec z}$ is unit vector along the rotation axis.
Pseudovacuum conditions for the magnetic field,
$
  \hat{\vec z}\times\vec{B}^\prime = 0,
$
are applied on the disk surfaces at
$
z = \pm H,
$
where prime signifies small disturbance. The conditions for the flow assume impenetrable and stress-free boundaries. 
To treat large radial distances we change to a new variable, 
\begin{equation}
  y = \frac{s/s_0}{1+ s/s_0}, \ \ \  0 \leq y \leq 1. 
  \label{6}
\end{equation}
and use uniform numerical grid in $y$. In effect, we did not compute up to $y=1$ but imposed a side boundary at $y = 0.99$ with the condition that all disturbances vanish at this boundary. 

The magnetic and velocity disturbances are expressed in terms of scalar potentials, e.g.
\begin{eqnarray}
   \vec{B}^\prime &=& \hat{\vec z}\times\vec{\nabla} B + 
   \vec{\nabla}\times\left(\hat{\vec z}\times\vec{\nabla}A\right) ,
   \label{7}
\end{eqnarray}
so that the disturbances are automatically divergence-free. The Fourier expansions in $z$ and in the azimuthal coordinate $\phi$ of the cylinder coordinate system, 
\begin{equation}
   B = \mathrm{e}^{\sigma t}\sum_{l,m}\mathrm{e}^{\mathrm{i}m\phi} \left(B_{lm}^\mathrm{S}(s)\cos\left(\pi(l -1/2)z\right)
   + B_{lm}^\mathrm{A}(s)\sin (\pi lz)\right) ,
   \label{8}
\end{equation}
are applied. Summation in Eq.\,(\ref{8}) runs over $l = 1,2,3,...$ and $m=0,1,2,...$. The mathematical treatment of the velocity and vorticity disturbances are quite similar.

The equation system for the disturbances split into a set of independent equations for different $l$ and $m$. Also the terms marked by the upper indeces S and A in (\ref{8}) are not mixed by the equations. The indices signify the magnetic modes symmetric and antisymmetric relative to the midplane of the disk. We shall use the notation Sm and Am for the symmetric and antisymmetric modes where $m$ is the azimuthal wave number. Note that Sm and Am represent families of modes that can be further distinguished by the vertical wave number $l$. For fixed $m$, $l$ and the symmetry type we have an  eigenvalue problem for ordinary differential equations in the variable $y$. The problem is solved numerically. 

The problem has three governing parameters, i.e. the magnetic Reynolds number (Rm), the Lundquist number (S),
\begin{equation}
   \mathrm{Rm} = \frac{\Omega_0 H^2}{\eta}, \ \ \ \ \ \ \ \ \ \ \ \ \ \
   \mathrm{S} = \frac{V_\mathrm{A} H}{\eta},
      \label{10}
\end{equation}
where $V_\mathrm{A} = B_0/\sqrt{\mu\rho}$ is the Alfv\'en velocity, and the magnetic Prandtl number
\begin{equation}
{\rm Pm}=\frac{\nu}{\eta} .
\label{10a}
\end{equation}
The Reynolds number Re = Rm/Pm and the Hartmann number Ha = S/$\sqrt{\mathrm{Pm}}$ were other possible choices. We shall see, however, that the parameters set (\ref{10}) is most convenient for MRI analysis for  small Pm.
\section{Marginal stability from local analysis}
For disturbances of small spatial scale compared to local radius, the differential rotation can be approximated by a plane shear flow which leads to the local approximation (Hawley \& Balbus \cite{HB91}). For the simplest case of plane-wave disturbances with  ${\vec k}\| \hat{\vec z}$ one finds the dispersion relation 
\begin{eqnarray}
  \lefteqn{\left(\sigma + \eta k^2\right)^2\left( \left(\sigma + \nu k^2\right)^2
   + 2\left(2 - q\right)\Omega^2\right)+}
   \nonumber \\
   &&\ \ \ \ \ + \Omega_{\rm A}^2\left( \Omega_{\rm A}^2 - 2 q \Omega^2
   + 2\left(\sigma + \nu k^2\right)\left(\sigma +\eta k^2\right)
   \right)
   = 0 ,
\label{11}
\end{eqnarray}
where $\sigma$ is the eigenvalue, $\Omega_\mathrm{A} = kV_\mathrm{A}$ is the magnetic frequency and $q = -\mathrm{d}\log\Omega /\mathrm{d}\log s$ is the local shear. An instability for the considered case can come only as a change of stability (Chandrasekhar \cite{C61}). Therefore, we can put $\sigma =0$ in Eq.\,(\ref{11}) to get the neutral stability equation defining the boundary between stability and instability (Kitchatinov \& R\"udiger \cite{KR04}): 
\begin{equation}
    \mathrm{Rm}^2 = \frac{\left(\mathrm{Pm} + \mathrm{S}^2\right)^2}
    {2\left( q\mathrm{S}^2 - 2 + q\right)} .
    \label{12}
\end{equation}
The dimensionless quantities (\ref{10}) are  redefined in terms of the wave number ($H \rightarrow k^{-1}$). 

\begin{figure}
   \centering
   \includegraphics[width=7.0cm]{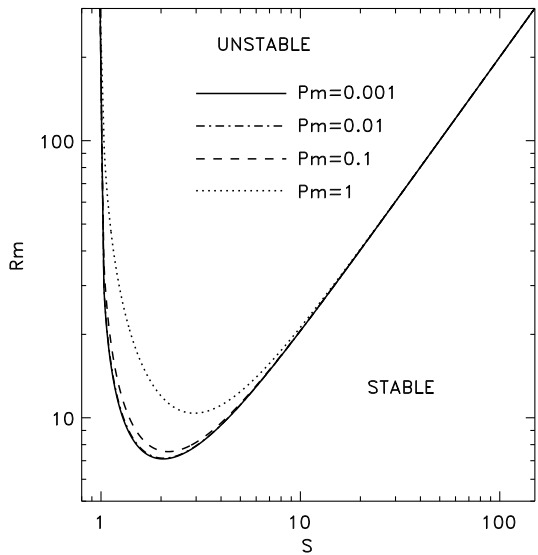}
   \caption{Neutral stability lines for most easily excited S0 modes and various Pm$\leq$1. The lines do not depend on Pm provided that Pm is small.
              }
   \label{f2}
\end{figure}

Equation (\ref{12}) shows that the instability requires sufficiently large Rm exceeding  
\begin{equation}
  \mathrm{Rm}_\mathrm{min} = \sqrt{\mathrm{Pm}\frac{2}{q} 
  \left( 1 + \frac{2-q}{q\mathrm{Pm}}\right)} ,
  \label{13}
\end{equation}
which corresponds to  $\mathrm{S} = \left( \mathrm{Pm} +2(2-q)/q\right)^{1/2}$. For  $\mathrm{Rm}\gg \mathrm{Rm}_\mathrm{min}$, the instability only exists for S between a lower and an upper limit, i.e.
$
  \sqrt{(2-q)/q} \leq \mathrm{S} \leq \sqrt{2q}\ \mathrm{Rm}. 
$
For small Pm the Eqs. (\ref{12}) and (\ref{13}) loose their dependence on Pm so that the viscosity completely drops off.  The instability in this limit is fully controlled by  Rm and S both defined in terms of the magnetic diffusivity. 
The computations of the most unstable axisymmetric modes of MRI for our global model    confirm these  results.  Figure\,\ref{f2} shows that the entire stability map in the Rm-S plane  does not depend on Pm  for small   Pm (R\"udiger \& Kitchatinov \cite{RK05}). Also the growth rates of the linear instability   for small Pm are fully controlled by Rm and S. 

\section{Nonaxisymmetric modes}
The magnetic Prandtl number is very small for cool protostellar and protoplanetary disks, it is also small for liquid metals of laboratory experiments but it is very large for galaxies (cf. Brandenburg \& Subramanian \cite{BS05}). Properties of MRI can be expected to vary strongly between the two cases of small and large Pm (see Lesur \& Longaretti \cite{LL07}; Fromang et al. \cite{Fea07}). The linear theory, however, shows that for $\rm Pm<1$ the viscosity does not play any role. This is in contradiction with Lesur \& Longaretti (2007).
\subsection{Small Pm}
\begin{figure}
   \centering
   \includegraphics[width=7.0cm]{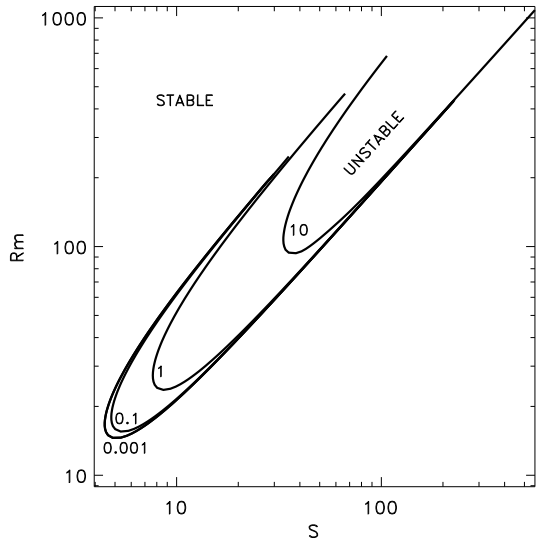}
   \caption{Neutral stability lines for nonaxisymmetric S1 modes of the lowest vertical wave number ($l = 1$) and various Pm. The lines are marked by the corresponding magnetic Prandtl numbers. 
              }
   \label{f3}
\end{figure}

\begin{figure}
   \centering
   \includegraphics[width=7.0cm]{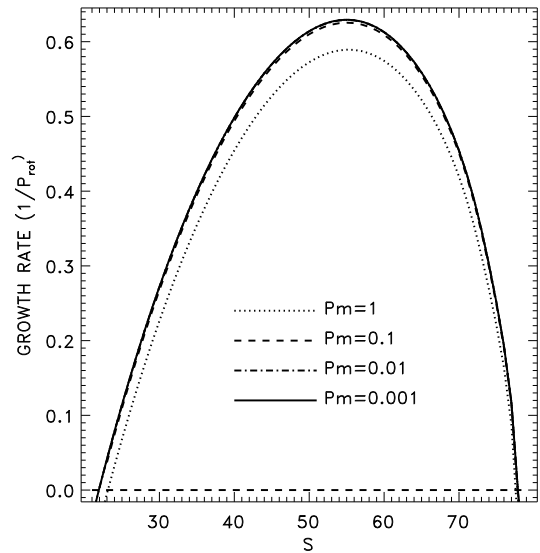}
   \caption{Growth rates of nonaxisymmetric S1 modes of the lowest vertical wave number ($l = 1$) for various Pm. The plot shows the growth rates in the range of magnetic fields producing the instability for ${\rm Rm}=150$. 
              }
   \label{f4}
\end{figure}

The results for the nonaxisymmetric modes confirm that MRI for small Pm is  not sensitive to viscosity. The neutral stability lines for $m=1$ given in  Fig.\,\ref{f3} approach one and the same contour as Pm decreases. The growth rates  show the same tendency (Fig.~\ref{f4}). Note that one cannot distinguish the cases of Pm = 0.01 and Pm = 0.001 by their growth rates or stability maps. 

This simple scaling should be important for numerical simulations. The magnetic Prandtl numbers in astrophysical bodies can be so small that they are not accessible for simulations. We have shown  with our linear theory that the basic instability for moderately small magnetic Prandtl number (${\rm Pm}\simeq 0.1$)  closely reproduce the results for indefinitely small  Pm (provided that the results are expressed in terms of Rm and S or other parameters not including the viscosity). The scaling means that MRI at small Pm does not develop so fine structure that the viscosity becomes important.

The strong-field limit of the instability domain in Fig.~\ref{f3} behaves like ${\rm Rm}\simeq 2$ S, i.e. the rotation is slightly superAlfv\'enic. The new feature in Fig.\,\ref{f3} is that the minimum field, $B_\mathrm{min}$, producing the instability also grows with Rm. This  positive slope is in contrast to the axisymmetric modes of Fig.\,\ref{f2} for which $B_\mathrm{min}$ approaches a small constant value as Rm grows. For given field strength, the nonaxisymmetric modes are switched off by too fast rotation. The relation ${\rm Rm}\simeq 10$ S provides an estimation for the upper limit on rotation rate for the S1 modes excitation. For ${\rm Rm}>10$\,S the S1 modes with $l=1$ no longer exist. With a plasma-$\beta$ with $\beta\simeq {\rm Rm}^2/{\rm S}^2$ one finds $\beta\lower.4ex\hbox{$\;\buildrel <\over{\scriptstyle\sim}\;$} 100$ as the instability condition for S1 modes with $l=1$. Hence, the magnetic field must be strong enough. We shall demonstrate below that for ${\rm Rm}>10$ S the realization of the mode stability  requires rather high numerical resolution in radial direction.  The nonaxisymmetric modes are necessary for self-sustained MRI-turbulence. A self-sustained turbulence in high Rm-regime may thus not be possible.  Another possibility is the nonaxisymmetric instability of an {\em imposed} azimuthal magnetic field (\lq\lq AMRI", R\"udiger et al. \cite{Rea07};  Simon \& Hawley \cite{SH09}). 

The increase of $B_\mathrm{min}$ with Rm for nonaxisymmetric MRI is a consequence of the winding effect of differential rotation. The pitch-angle of unstable disturbances near $B_\mathrm{min}$ is small (Kitchatinov \& R\"udiger \cite{KR04}), i.e. the winding is strong. The differential rotation converts azimuthal inhomogeneity of the nonaxisymmetric modes into a fine radial structure which is finally destroyed by diffusion.

The increase of $B_\mathrm{min}$ with Rm also appears for large Pm. In this case it is  reasonable to use  Re and S/Pm  to parameterize the rotation and the background field. The lines  show little dependence on large Pm when plotted in the plane of these parameters which now do not depend on the magnetic diffusion $\eta$.

\subsection{Overtones}
Another new feature of the nonaxisymmetric instability is that modes with higher vertical wave number ($l > 1$) are preferred for some parameter domains. For axial symmetry, the region of parameters where the S0 mode with $l=1$ is unstable includes the instability regions of all other modes.

\begin{figure}
   \centering
   \includegraphics[width=7.0cm]{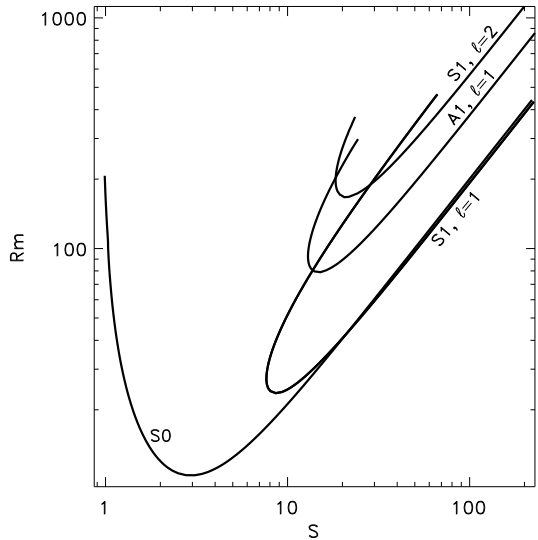}
   \caption{Common stability map of S0 mode ($l=1$) and several nonaxisymmetric modes of different structure in $z$-direction. S1 mode is preferred on the strong-field side of the plot. On the weak-field side, the neutral stability lines of nonaxisymmetric modes intersect. Pm =1.  
              }
   \label{f6}
\end{figure}

Figure\,\ref{f6} shows the neutral stability lines for several nonaxisymmetric modes together with the line for the most unstable axisymmetric S0 mode. The lines for the nonaxisymmetric modes intersect so that the modes with finer vertical structure are preferred on the weak-field side of the stability map. This is again related to the winding effect of differential rotation. The modes with finer vertical structure have larger Lorentz force to resist the winding of weak fields. Note that only the S1 mode with  $l=1$ can compete with S0 mode. There is a narrow region on the strong-field side of Fig.\,\ref{f6} where the mode S0  is stable but S1 not.

\begin{figure}
   \centering{
   \includegraphics[width=4.2cm]{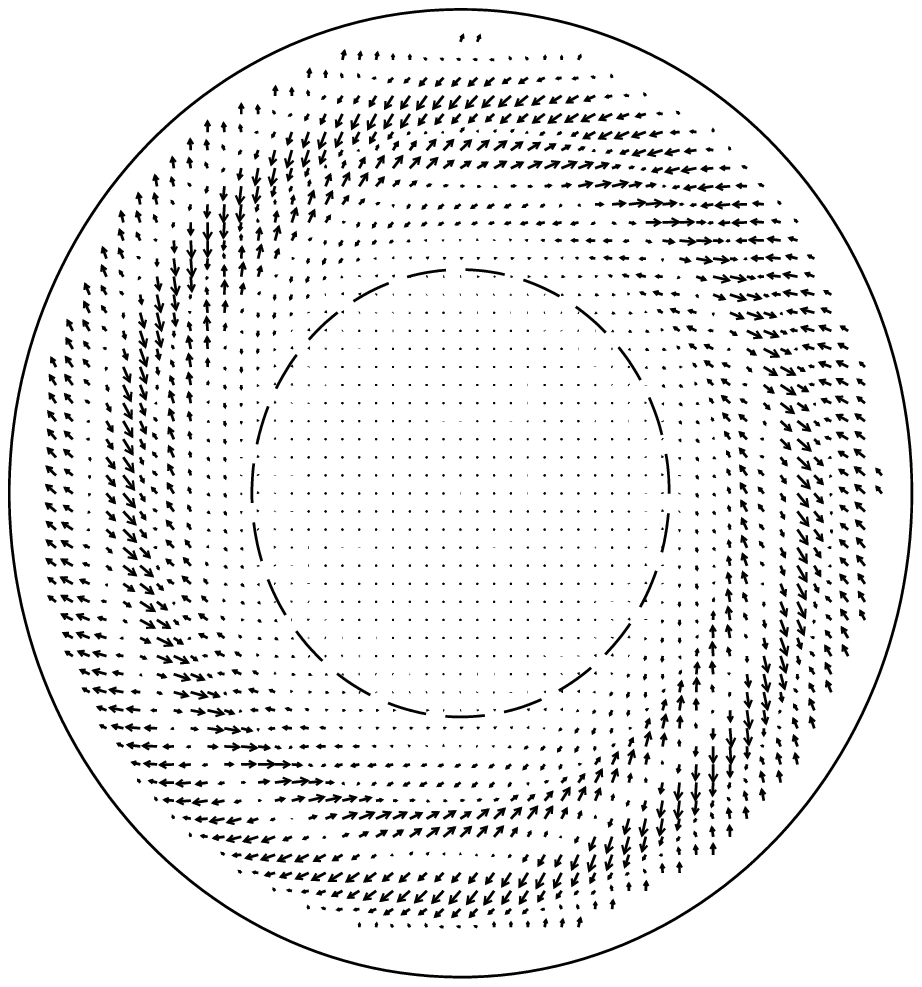}
   \hspace{0.1truecm}
   \includegraphics[width=4.2cm]{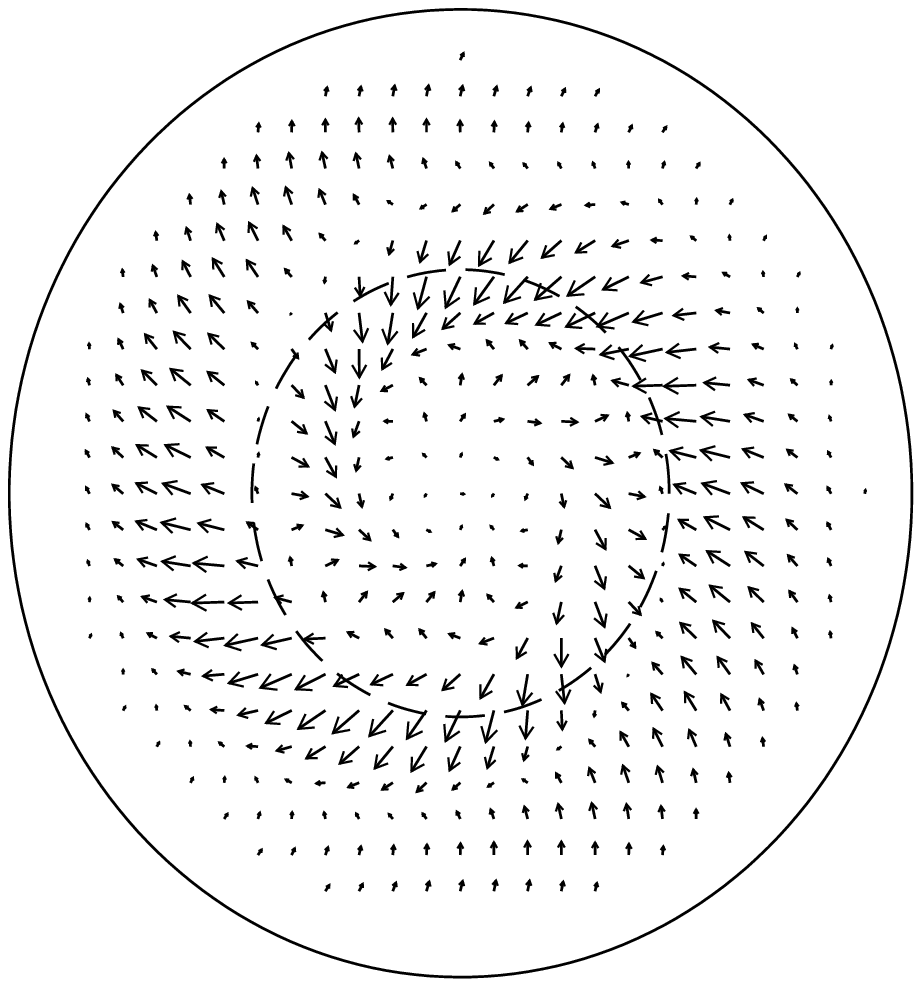}}
   \caption{Vectorplots of the unstable S1 modes ($l=1$) for the mid-plane of the disk after computations for ${\rm Rm}=100$ and ${\rm Pm}=1$. The left panel is for the field slightly exceeding $B_\mathrm{min}$ and the right panel - slightly below $B_\mathrm{max}$. The dashed circle shows the corotation radius $s = s_0$.
              }
   \label{f7}
\end{figure}

The positive slope on the upper ${\rm Rm}({\rm S})$ curve of the stability map also exists if higher $l$ modes are included. The existence of nonaxisymmetric magnetic instability is necessary for any form of MRI-dynamo. The preference of nonaxisymmetry for strong fields is promising for the dynamo and the self-sustained turbulence. The field strength should increase sufficiently fast with Rm, or the plasma-$\beta$  of shearing box simulations (Fromang et al. \cite{Fea07}) should not be too high, in order to probe this possibility. Otherwise, for too large $\beta$, i.e. for too weak fields the nonaxisymmetric modes do not appear. 
\section{The resolution problem}
The upper branches of the neutral stability lines for the nonaxisymmetric modes in Fig.\,\ref{f6} are always terminated because of   numerical resolution problem. The needed resolution  to follow the lines for higher Rm rapidly increases. Again the shearing by differential rotation is the reason. Figure\,\ref{f7} gives  vector plots of  unstable nonaxisymmetric magnetic modes for weak (close to $B_\mathrm{min}$) and strong (close to $B_\mathrm{max}$) background fields. Obviously, the disturbances in the  strong-field case resist the shearing by the differential rotation. On the other hand, the shearing of  weak fields produces  tightly wound spirals with increasing requirements for the numerical resolution. 

\begin{figure}
   \centering{
   \includegraphics[width=4.1cm]{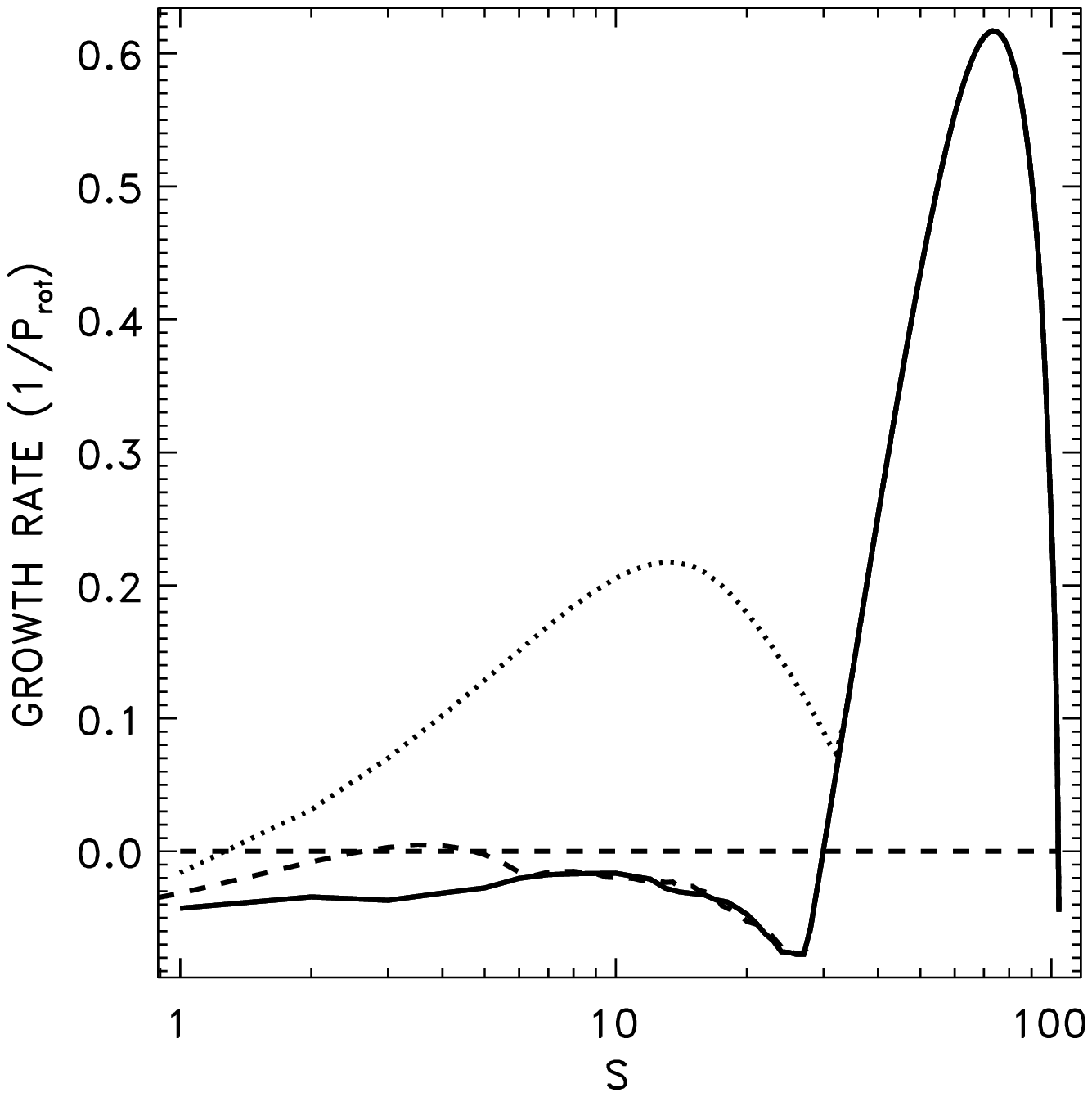}
   \hspace{0.1truecm}
   \includegraphics[width=4.1cm]{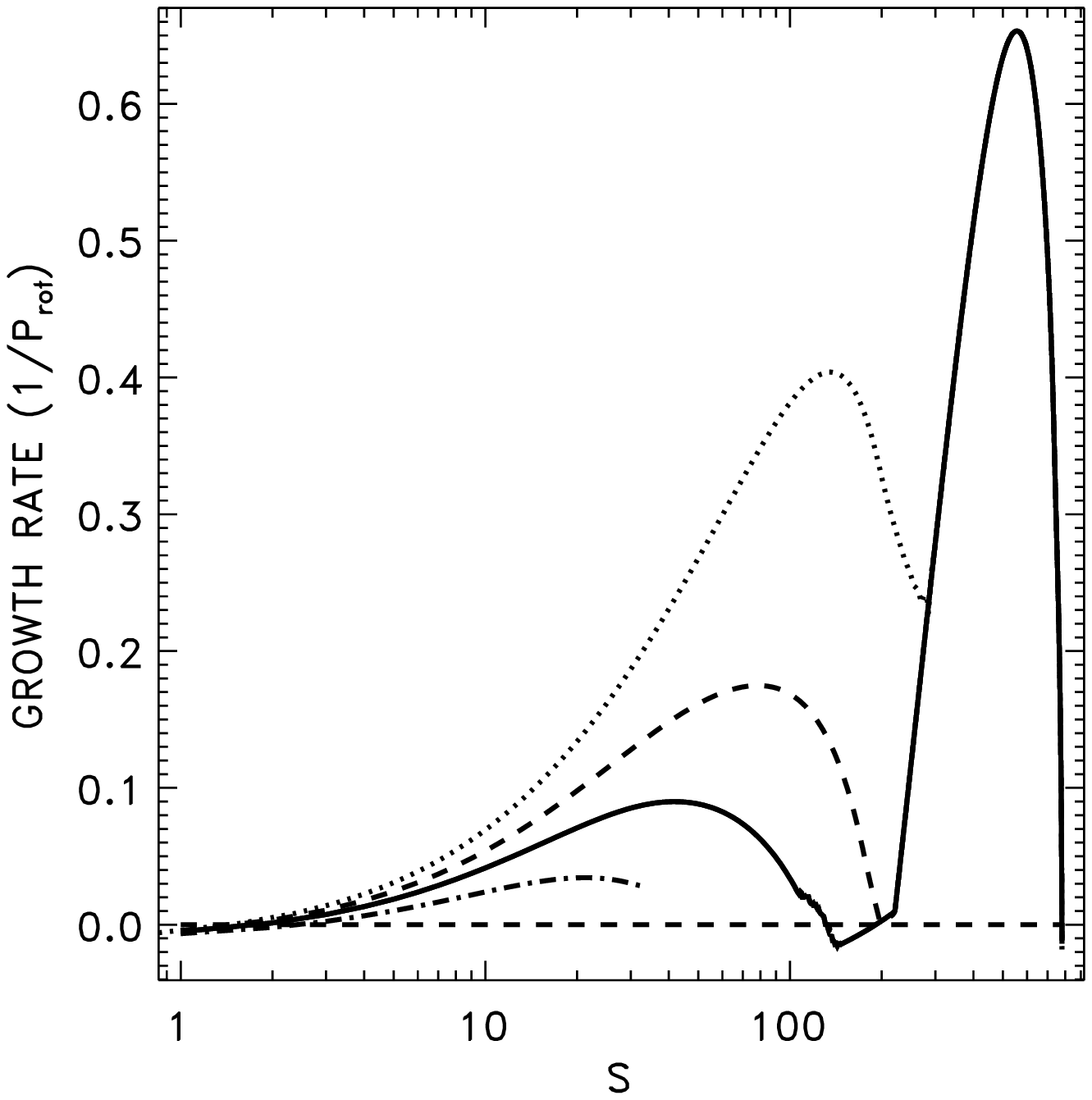}}
   \caption{Growth rates of S1 modes for ${\rm Re}=150$, ${\rm Pm}=1$ (left), and ${\rm Pm}=10$ (right) computed with different resolutions. The dotted, dashed, full, and dashed-dotted lines were obtained with 100, 150, 200, and 300 grid points in radius, respectively. 
              }
   \label{f8}
\end{figure}

In Fig.\,\ref{f8} the MRI growth rates calculated for the S1 mode computed with different numbers of radial grid points are shown. There is no resolution problem at all for strong fields. For weak fields, however, a too low  resolution produces an unreal  instability. This numerical artifact can be suppressed by increasing resolution. For fixed Reynolds number it is even  harder to do it for large Pm. In the given examples always more than 200 gridpoints are necessary to overcome the numerical artefacts. 

The resolution problem has also been  met in nonlinear shearing box simulations of MRI (Fromang \& Papaloizou \cite{FP07}). If our interpretation of the problem as a result of rotational shearing is correct then an increase of resolution only in one radial direction is necessary to resolve the problem.  Also their value of the plasma-$\beta$ of order 400 
leading to the relation $\rm Rm \simeq 20$ S appears to imply too weak magnetic fields outside of the region of nonaxisymmetric MRI. 

We conclude that a self-sustained MRI-turbulence may only be found with sufficiently {\em strong initial field}. The minimum field for the turbulence exceeds by at least one order of magnitude the minimum external field required for axisymmetric MRI and it increases with magnetic Reynolds number. 
\begin{acknowledgements}
This work was supported by the Alexander von Humboldt Foundation and by the Russian Foundation for Basic Research (project 09-02-91338).
\end{acknowledgements}


\begin{thebibliography}{}
\bibitem[1998]{BH98}
    Balbus,\,S.\,A., \& Hawley,\,J.\,F. 
    1998, Rev. Mod. Phys., 70, 1
\bibitem[2005]{BS05}
    Brandenburg,\,A., \& Subramanian,\,K.
    2005, Phys.\,Rep., 417, 1
\bibitem[1995]{BNST95}
    Brandenburg,\,A., Nordlund,\,\AA., Stein,\,R.\,F., \& Torkelsson,\,U. 
    1995, \apj, 446, 746
\bibitem[1961]{C61}
    Chandrasekhar,\,S.
    1961, Hydrodynamic and Hydromagnetic Stability
    (Oxford: Clarendon Press)
\bibitem[1946]{E46}
    Elsasser,\,W.\,M. 
    1946, PhRv, 69, 106
\bibitem[2007]{FP07}
    Fromang,\,S., \& Papaloizou,\,J.
    2007, \aap, 476, 1113
\bibitem[2007]{Fea07}
    Fromang,\,S., Papaloizou,\,J., Lesur,\,G., \& Heinemann,\,T.
    2007, \aap, 476, 1123
\bibitem[1991]{HB91}
    Hawley,\,J.\,F., \& Balbus,\,S.\,A.
    1991, \apj, 376, 223
\bibitem[1996]{HGB96}
    Hawley,\,J.\,F., Gammie,\,C.\,F., \& Balbus,\,S.\,A.  
    1996, \apj, 464, 690
\bibitem[1997]{KM97}
    Kitchatinov,\,L.\,L., \& Mazur,\,M.\,V. 
    1997, \aap, 324, 821
\bibitem[2004]{KR04}
    Kitchatinov,\,L.\,L., \& R\"udiger,\,G. 
    2004, \aap, 424, 565
\bibitem[2005]{LL07}
    Lesur,\,G., \& Longaretti,\,P.-Y.
    2007, \mnras, 378, 1471
\bibitem[2005]{RK05}
    R\"udiger,\,G., \& Kitchatinov,\,L.\,L.
    2005, \aap, 434, 629
\bibitem[2007]{Rea07}
    R\"udiger,\,G., Hollerbach,\,R., Schultz,\,M., \& Elstner,\,D. 
    2007, \mnras , 377, 1481
\bibitem[2009]{SH09}
    Simon,\,J.\,B., \& Hawley,\,J.\,F. 
    2009, arXiv:0906.5352
\end{thebibliography}
\end{document}